\documentclass[prd,preprint,nofootinbib,a4paper,showpacs]{revtex4}

\usepackage{amsmath,amssymb}
\usepackage{graphicx}

\newcommand{\mcl}{\mathcal{L}}
\newcommand{\mcn}{\mathcal{N}}
\newcommand{\mco}{\mathcal{O}}
\newcommand{\mcs}{\mathcal{S}}
\newcommand{\rt}{\sqrt{2}}

\newcommand{\rtkap}{\rt\,\kappa}
\newcommand{\rtkapinv}{\frac{1}{\rtkap}}
\newcommand{\rtkapsqinv}{\frac{1}{\rt\,\kappa^2}}

\newcommand{\auxfun}{f}

\DeclareMathOperator{\arctanh}{arctanh}
\DeclareMathOperator{\arcsinh}{arcsinh}

\newcommand{\half}{\frac{1}{2}}

\begin{document}


\title{Survival of scalar zero modes in warped extra dimensions}

\author{Damien P. George}
\email{dpgeorge@nikhef.nl}
\affiliation{Nikhef Theory Group, Science Park 105, 1098 XG Amsterdam, The Netherlands}

\date{\today}
\preprint{NIKHEF/2011-003}

\begin{abstract}
Models with an extra dimension generally contain background scalar
fields in a non-trivial configuration, whose stability must be
ensured.  With gravity present, the extra dimension is warped by
the scalars, and the spin-0 degrees of freedom in the metric mix
with the scalar perturbations.  Where possible, we formally solve
the coupled Schr\"odinger equations for the zero modes of these
spin-0 perturbations.  When specialising to the case of two
scalars with a potential generated by a superpotential, we are
able to fully solve the system.  We show how these zero modes can
be used to construct a solution matrix, whose eigenvalues tell
whether a normalisable zero mode exists, and how many negative
mass modes exist.  These facts are crucial in determining
stability of the corresponding background configuration.  We
provide examples of the general analysis for domain-wall models of
an infinite extra dimension and domain-wall soft-wall models.  For
5D models with two scalars constructed using a superpotential, we
show that a normalisable zero mode survives, even in the presence
of warped gravity.  Such models, which are widely used in the
literature, are therefore phenomenologically unacceptable.
\end{abstract}

\maketitle


\section{Introduction}

A plausible way to extend the standard model is to embed it in
one or more extra dimensions.  This opens up a new set of
model-building tools which can help to solve a diverse range of
theoretical and phenomenological problems, as well as yielding
distinct collider signatures such as Kaluza-Klein (KK) modes.
Almost all extra-dimensional models require one or more background
scalar fields in some non-trivial configuration.  For example, to
generate a domain-wall which localises chiral
fermions~\cite{Rubakov:1983bb}, to stabilise the size of a compact
extra dimension~\cite{Goldberger:1999uk,Csaki:2000zn}, to
generalise the Randall-Sundrum warped-space~\cite{Randall:1999vf}
to a smoothed-out version~\cite{Csaki:2000fc}, or to cut off the
extra dimension at a singularity~\cite{Gremm:2000dj,Karch:2006pv}.
Domain-wall models, whether they have an
infinite~\cite{Davies:2007xr} or compact~\cite{Aybat:2010sn} extra
dimension, make heavy use of background scalar configurations
as a field-theoretic substitution for fundamental branes.

Given the ubiquity and necessity of background scalar fields, it
is important to understand both their statics and dynamics.  The
problem is best thought about in terms of a ground state, upon
which exist perturbations.  We want to know which particular
scalar configurations have the lowest energy and are stable, and
what the perturbations about a background lead to in terms of
effective 4D modes.  These two issues are closely related.  The
existence of negative-mass modes (tachyonic KK modes) signals
an instability of the corresponding background.  Massless modes
may also signal an instability~\cite{Aybat:2010sn}, or be harmless
in the case of a translation mode.  Positive-mass modes always
exist and their precise spectrum is what distinguishes these
extra-dimensional models at a particle collider.

For the case of a single scalar field in a flat compact extra
dimension, a general method for determining the lowest energy
configuration has been worked
out~\cite{Toharia:2007xe, Toharia:2007xf}.  The inclusion of
gravity in the analysis presents some complications because of the
coupling of the scalar fields to gravity.  This coupling
generically warps the extra dimension~\cite{Csaki:2000fc} and the
scalar perturbations mix with the spin-0 degrees of freedom in the
metric.  For this warped case with one extra dimension there have
been some general stability analyses with a single background
scalar~\cite{Kobayashi:2001jd, Toharia:2008ug}, and some initial
work on the multiple scalar
case~\cite{Toharia:2010ex, Aybat:2010sn}.
For the case of a single scalar in multiple extra dimensions it
has also been shown that the scalar and metric spin-0 modes
mix~\cite{Underwood:2010pm}.
Related analyses determining the spin-0 spectrum of multiple
scalars in 5D have been done in the context of the AdS/CFT
correspondence~\cite{Berg:2005pd, Berg:2006xy, Elander:2009bm}.
In particular, an algorithm for computing the scalar spectrum in
general 5D compact models has been
prescribed~\cite{Elander:2010wd}.

Despite the complications introduced by gravity, it is still
possible to find the effective coupled Schr\"odinger equations
which describe the KK modes of multiple scalars in a warped
background~\cite{Aybat:2010sn}.  It is the aim of the current
paper to, when possible, formally solve this set of Schr\"odinger
equations for the massless KK modes~--- the zero modes~--- in the
case of a 5D bulk with no gravity (flat) and with gravity
(warped).  In addition to providing closed form solutions for the
zero modes for a large number of cases, we shall also discuss how
these solutions can be used to determine if any normalisable zero
modes exist and whether or not the background is perturbatively
stable.  Some relevant examples shall be provided.

One reason for studying the zero modes of the system is that, if
they exist, they should play a large role at low energies in the
effective 4D theory.  For example, zero modes of a 5D fermion are
generally used to implement the fermions of the standard
model~\cite{Davies:2007xr}.  When constructing domain-wall models
using scalar background fields in flat space, one always obtains a
spin-0 zero mode corresponding to the broken translation symmetry.
This degree of freedom is not welcome in the effective 4D theory
since we have never observed such a particle, which would manifest
as a ``fifth force''.  Adding warped gravity can cure this
problem since this removes the translation zero mode, as shown by
Shaposhnikov et.\ al.~\cite{Shaposhnikov:2005hc} for the case of a
single background scalar.  In this paper this result is extended
to the case of multiple background scalar fields: the zero mode of
translation does not survive, no matter how many scalars.  It is
possible though, and we shall give some explicit examples, that
additional scalars introduce additional zero modes which do
survive in the presence of gravity.  Our examples of such models
are constructed using the superpotential approach, and we argue
that these models are phenomenologically unacceptable.

The paper is organised as follows.  In Section~\ref{sec:zm} we
first present the zero mode solutions, for the flat and warped
cases, with general potential $V$ and also specialising to a
superpotential $W$ with $N=2$ scalars.  For this latter case we
give all four independent zero modes in closed analytic form.
Section~\ref{sec:use-of-zm} discusses the construction of a
solution matrix, and how its eigenvalues can be used to find
normalisable zero modes, and count the number of normalisable
negative modes.  Following this we look in Sections~\ref{sec:dw}
and~\ref{sec:sw} at specific domain-wall models in both an
infinite and a compact extra dimension, and show that zero modes
can survive in the presence of gravity.  We conclude in
Section~\ref{sec:concl}.  Appendix~\ref{sec:app} summarises the
method of reduction of order for ordinary differential equations.


\section{The zero mode solutions}
\label{sec:zm}

In this section we study spin-0 perturbations of $N$ real scalar
fields in a 5D bulk for both a flat and warped extra dimension.
The scalar fields are assumed to have some non-trivial background
profile along the extra dimension, such as a kink.  The aim is to
derive formal solutions for the extra-dimensional profile of the
zero modes, that is, the massless perturbations around the
background.  We shall concentrate mainly on the $N=2$ case, and,
for part of the analysis, specialise to potentials $V$ that are
generated by a superpotential $W$.

Throughout this paper we work with the matter Lagrangian
\begin{equation}
  \label{eq:lag}
  \mcl_m = -\frac{1}{2}g^{MN}\partial_M\Phi_i\partial_N\Phi_i - V(\Phi_i) \:,
\end{equation}
where $g_{MN}$ is the 5D metric with signature $(-++++)$,
$\Phi_i(x^\mu,y)$ are $N$ scalar fields indexed by $i=1\ldots N$,
$x^\mu$ is the 4D sub-spacetime, $y$ is the coordinate of the
extra dimension, and $V(\Phi_i)$ is the scalar potential.
Repeated scalar field indices are always summed over.
Perturbations of the scalar fields around some arbitrary
background are written as
\begin{equation}
  \label{eq:scalar-pert}
  \Phi_i(x^\mu,y) = \phi_i(y) + \varphi_i(x^\mu,y) \:.
\end{equation}
The background profiles $\phi_i(y)$ depend only on the extra
dimension, while the perturbations are general functions of all
spacetime coordinates, and are required to be relatively small:
$\varphi_i \ll \phi_i$.

From the equations of motion for the scalars one can obtain
$N$ coupled, one dimensional, time-independent Schr\"odinger-like
equations for the perturbations.\footnote{In the case with gravity,
we present $N+1$ Schr\"odinger equations plus one constraint
equation, yielding effectively $N$ Schr\"odinger equations.}
The independent variable of the these equations is the extra
dimension $y$ and the eigenvalue corresponds to the mass of the
KK mode of the perturbation.  Where possible, we shall
formally solve this system of coupled Schr\"odinger equations for
the case of a zero eigenvalue.  Since we have $N$ second order,
linear, ordinary differential equations (ODEs) we expect to obtain $2N$
linearly independent solutions.  These $2N$ solutions are not
necessarily physical (that is, are not normalisable), but they
do form a basis from which one can construct the unique solution
for any given initial, boundary and/or normalisability conditions.
The zero mode solutions are useful for studying the perturbative
stability of the background configuration, as shall be discussed
in Section~\ref{sec:use-of-zm}.

\subsection{Flat case}

We analyse first the gravity-free, flat-space scenario where the
action is given simply by $\mcs=\int \mcl_m \, d^4x \, dy$.
The discussion is divided into the case of a general $V$, and the
case where $V$ is generated by a (fake) superpotential $W$.

\subsubsection{General $V$}

In flat space, the general background equations are
\begin{equation}
  \label{eq:flatv-bg}
  \phi_i'' - V_i = 0 \:,
\end{equation}
and the coupled Schr\"odinger-like equations for the perturbations
are
\begin{equation}
  \label{eq:flatv-pert}
  -\varphi_i'' + V_{ij} \varphi_j = \Box \varphi_i \:.
\end{equation}
Prime denotes derivative with respect to $y$, subscripts $i$ and
$j$ on $V$ denote a derivative with respect to $\Phi_i$ and
$\Phi_j$, and $\Box=\partial^\mu\partial_\mu$.  In addition,
$V_{ij}$, which is a function of $\Phi_i$, must be evaluated
on the background solution, $\Phi_i=\phi_i(y)$, where $\phi_i(y)$
is a solution to equation~\eqref{eq:flatv-bg}.  As usual,
separation of variables of $x^\mu$ and $y$ is the correct way to
proceed here.  For the sake of reducing the number of field
variables we shall abuse notation slightly by using
$\varphi_i$ to denote both the full perturbation which is a
function of $x^\mu$ and $y$, as well as the separated factor that
depends only on $y$.  Separation of variables then proceeds as per
$\varphi_i(x^\mu,y)=\varphi_i(y)\rho(x^\mu)$, with $\rho$ the 4D
KK mode with mass $m$, such that $\Box\rho=m^2\rho$.

One now solves equation~\eqref{eq:flatv-pert} for the profiles
$\varphi_i(y)$ and corresponding allowed mass values, obtaining
a tower of modes.  Then one takes the original 5D action and
substitutes $\Phi_i=\phi_i(y)+\varphi_i(y)\rho(x^\mu)$ with
$\varphi_i(y)$ a particular solution.  The extra dimension can
then be integrated out, leaving an effective 4D action for the
mode $\rho$.  To second order in $\rho$, this action is
\begin{equation}
  \mcs = \int d^4x \left[
    - \varepsilon_\text{bg}
    + \mcn \left(
      - \half \partial^\mu \rho \partial_\mu \rho
      - \half m^2 \rho^2 \right) \right]
    + \text{(surface terms)} \:.
\end{equation}
Here, $\varepsilon_\text{bg}$ is the energy density of the
scalar background configuration.  The normalisation constant for
the mode is $\mcn=\int\varphi_i^2dy$ and, so long as we pick
a solution $\varphi_i$ that is normalisable, one can scale said
solution to obtain $\mcn=1$.  The action then describes a
canonical, 4D scalar field.  The surface terms in the action
are of the form $\int S' d^4x \, dy$, where $S$ is one of
\begin{subequations}
\begin{align}
  S_1 &= -\phi_i' \varphi_i \:,\\
  S_2 &= -\half \varphi_i \varphi_i' \:.
\end{align}
\end{subequations}
The requirements that these terms independently vanish on the
boundaries of the extra dimension, and that $\mcn$ is finite, pick
out the physical modes of the KK tower.

As mentioned previously, we are interested in the zero modes, and
shall look for formal solutions to equation~\eqref{eq:flatv-pert}
when $\Box\varphi_i=0$.  For general $V$ and all $N$ one solution
is\footnote{Equation~\eqref{eq:flatv-pert} is linear in
$\varphi_i$ and we are free to scale any solution by an arbitrary
constant, a constant which in some cases is dimensionful.  For
brevity, and because we are providing just formal solutions, we
leave this constant out.  Hence the units of this equation, and
some of the equations that follow, do not match.}
\begin{equation}
  \label{eq:flatv-zm1}
  \varphi_i^{(1)} = \phi_i' \:,
\end{equation}
where $\phi_i$ is the background scalar field configuration.
Note that this solution is actually a vector of length $N$.  It
is the well-known translation mode of the background, since it is
the first term in a Taylor expansion of the shifted background:
$\phi_i(y+\epsilon)=\phi_i(y)+\epsilon\phi_i'(y)+\mco(\epsilon^2)$.
For the case of $N=1$, where there are $2N=2$ independent
solutions, we can use the $\varphi_i^{(1)}$ solution to perform
reduction of order (see Appendix~\ref{sec:app}) and obtain the
other solution:
\begin{equation}
  \label{eq:flatv-zm2-n1}
  \varphi_i^{(2,N=1)} = \phi' \int \left(\phi'\right)^{-2} dy \:.
\end{equation}
For $N>1$ we could also perform reduction of order, but we would
still have a relatively large (at least third order for $N=2$) ODE
to solve.  At this point we shall be content with having only the
translation solution for general $N$.

\subsubsection{$V$ generated by a superpotential}

It is possible to make progress and obtain an additional zero mode
solution when the form of $V$ is restricted to
\begin{equation}
  \label{eq:flatw-pot}
  V = \frac{1}{2} W_i^2 \:,
\end{equation}
where $W(\Phi_i)$ is a fake superpotential (it is just a formal
construction and does not indicate a supersymmetry).  As it does
on $V$, a subscript $i$ on $W$ denotes a derivative with respect
to $\Phi_i$.  In this case the background equations can be
written as
\begin{equation}
  \phi_i' = W_i \:,
\end{equation}
and the perturbation equation~\eqref{eq:flatv-pert} becomes
\begin{equation}
  -\varphi_i'' +  (W_{ij} W_{jk} + W_{ijk} W_j) \varphi_k = \Box \varphi_i \:.
\end{equation}
Here, $W_{ij}$ are to be evaluated on the background solution, and
then they become functions of $y$.  The utility of the
superpotential approach comes from the fact that this perturbation
equation can be factorised as~\cite{Bazeia:2010yp}
\begin{equation}
  \label{eq:flatw-pert}
  (\partial_y \delta_{ij} + W_{ij})
    (-\partial_y \delta_{jk} + W_{jk})\varphi_k = \Box \varphi_i \:.
\end{equation}
Now when we look for zero modes, half of the solutions can be
obtained by solving the much simpler equation
\begin{equation}
  \label{eq:flatw-pert-half}
  (-\partial_y \delta_{ij} + W_{ij})\varphi_j = 0 \:.
\end{equation}

For all $N$ the translation mode still exists,
\begin{equation}
  \label{eq:flatw-zm1}
  \varphi_i^{(1)} = W_i \:,
\end{equation}
and for $N=1$ the second solution will be given by
equation~\eqref{eq:flatv-zm2-n1}.  For $N=2$ the situation becomes
more interesting than the general $V$ case.  We can use the
$\varphi_i^{(1)}$ solution to reduce the order of
equation~\eqref{eq:flatw-pert-half} from two to one, and then
solve the resulting first order ODE to obtain a second zero
mode solution:
\begin{equation}
  \label{eq:flatw-zm2}
  \varphi_i^{(2)} = W_i \int \frac{JZ}{X^2} \, dy + a_i\frac{J}{X} \:.
\end{equation}
Here we have defined
\begin{subequations}
\begin{align}
  J &= \exp \left[ \int (W_{11}+W_{22}) dy \right] \:,\\
  X &= a_2 W_1 - a_1 W_2 \:,\\
  Z &= a_1 a_2 (W_11 - W_22) - (a_1^2-a_2^2) W_12 \:.
\end{align}
\end{subequations}
$a_1$ and $a_2$ are constants and must be chosen such that the
entity $X$ (which is a function of $y$) is non-zero throughout the
entire domain of $y$.  For example, for systems where $\phi_1$ is
odd and $\phi_2$ is even one can choose $a_1=0$, $a_2=1$.  We
stress that different values of the $a_i$ do not generate
independent zero mode solutions.

At this point we need to make a few remarks about integration
constants.  There are two integrals in the solution
$\varphi_i^{(2)}$.  The constant coming from the integral in $J$
yields an overall normalisation factor for the zero mode
solution.  The constant in the integral in the first term in
equation~\eqref{eq:flatw-zm2} pulls out a constant multiple of
$\varphi_i^{(1)}$, which effectively adds a multiple of this other
zero mode solution.  Thus our two integration constants amount to
taking linear combinations of two linearly independent zero mode
solutions.  Alternatively, one can fix these constants of
integration to zero and take linear combinations of
equations~\eqref{eq:flatw-zm1} and~\eqref{eq:flatw-zm2}.  Either
way, we have a closed form for the general solution to
equation~\eqref{eq:flatw-pert-half} when $N=2$.

There are two more linearly independent zero mode solutions for
the $N=2$ case.  We cannot obtain them in closed form like the
first two, but we can make some progress.  Using the two known
solutions $\varphi_i^{(1)}$ and $\varphi_i^{(2)}$ the fourth
order equation~\eqref{eq:flatw-pert} can be reduced to a second
order ODE.  This allows us to write the third and fourth zero
mode solutions as
\begin{equation}
  \label{eq:flatw-zm34}
  \varphi_i^\text{(3,4)} = A_1 \varphi_i^{(1)} + A_2 \varphi_i^{(2)} \:,
\end{equation}
where
\begin{subequations}
\begin{align}
  A_1 &= \int \frac{1}{J} \left( w_3 \varphi_2^{(2)} - w_4 \varphi_1^{(2)} \right) dy \:,\\
  A_2 &= \int \frac{1}{J} \left( -w_3 \varphi_2^{(1)} + w_4 \varphi_1^{(1)} \right) dy \:,\\
  w_3' &= -W_{11} w_3 - W_{12} w_4 \:,\\
  w_4' &= -W_{12} w_3 - W_{22} w_4 \:.
\end{align}
\end{subequations}
The equations for $w_3(y)$ and $w_4(y)$ constitute the second
order ODE which can be solved only with specific information about
$W$.  Since it is second order, there will be two sets of
solutions for the pair $\{w_3,w_4\}$, and substituting these
solutions into the equations for $A_1$ and $A_2$ will yield,
through equation~\eqref{eq:flatw-zm34}, the final two independent
zero modes.  There are four integration constants in the above
system of equations, as expected.  Those in $A_1$ and $A_2$ add,
respectively, a constant multiple of $\varphi_i^{(1)}$ and
$\varphi_i^{(2)}$ to $\varphi_i^{(3,4)}$.  The other two come from
solving for $w_3$ and $w_4$.  (The constant from $J$ can be
absorbed in a rescaling of $w_3$ and $w_4$.)

\subsection{Warped case}

We now repeat the previous gravity-free calculation for the case
with gravity.  It turns out that the Einstein constraint equation
allows one to obtain additional zero mode solutions.

The 5D action for $N$ scalar fields coupled minimally to gravity
is
\begin{equation}
  \label{eq:warp-act}
  \mcs = \int d^4x \, dy \sqrt{-g} \left( \frac{1}{6\kappa^2} R + \mcl_m \right) \:,
\end{equation}
where $\kappa^2 = 1/6M^3$ and $M$ is the 5D Planck mass.
Einstein's equations arising from this action are
$G_{MN}=3\kappa^2T_{MN}$ where the stress energy tensor is
$T_{MN}=\partial_M\Phi\partial_N\Phi + g_{MN} \mcl_m$.
We restrict our analysis to a warped metric ansatz, which is
actually the most general 5D metric that respects 4D Poincar\'e
invariance, and is used extensively in realistic models.  As for
perturbations of the metric, we need only consider scalar
perturbations, as vector and tensor perturbations decouple from
the spin-0 sector~\cite{Csaki:2000zn}.  With scalar perturbations
$F(x^\mu,y)$, the metric ansatz is
\begin{equation}
  \label{eq:warp-metric}
  ds^2 = e^{-2\sigma} (1-2F)\,\eta_{\mu\nu}\,dx^\mu dx^\nu + (1+4F)\,dy^2 \:.
\end{equation}
The warp factor exponent is $\sigma(y)$ and $\eta_{\mu\nu}$ is the
4D Minkowski metric.  For consistency of small perturbations we
require $F \ll 1$.  The perturbations of the scalar fields are as
in the previous section, equation~\eqref{eq:scalar-pert}.

We now look for formal zero modes of this set-up, first in the
case of a general scalar potential $V$, then in the case of $V$
generated by a superpotential $W$.

\subsubsection{General V}

With a general potential $V$ the background fields $\phi_i(y)$ and
$\sigma(y)$ satisfy the equations
\begin{subequations}
\begin{align}
  & \sigma'^2 = \frac{\kappa^2}{2} \left( \frac{1}{2} \phi_i'^2 - V \right) \:,\\
  & \phi_i'' - 4 \sigma' \phi_i' - V_i = 0 \:.
\end{align}
\end{subequations}
By $x^\mu$ scaling invariance, we are free to choose
$\sigma(0)=0$, leaving $2N$ integration constants for this set of
equations.  One of the redundant Einstein's equations, which is
sometimes useful, is $\sigma''=\kappa^2\phi_i'^2$.  For the
rest of this section, $\sigma$ and $\phi_i$ will be used to
denote solutions to these background equations, and the potential
$V$ and its derivatives with respect to $\Phi_i$ are to be
evaluated on this background.

For the perturbation equations, it is best to work with the new
variables $\chi(x^\mu,y)$ and $\psi_i(x^\mu,y)$ defined by
\begin{subequations}
\begin{align}
  F &= \frac{\kappa}{\rt} e^{2\sigma} \chi \:,\\
  \label{eq:varphi-psi-relation}
  \varphi_i &= e^{2\sigma} \psi_i \:.
\end{align}
\end{subequations}
We shall write these $N+1$ components as a vector
$\Psi = (\chi, \psi_i)$ when it makes things neater; $\Psi_m$ is
an indexed version with $m=0\ldots N$ so that $\Psi_0=\chi$ and
$\Psi_i=\psi_i$.  In physical $y$ coordinates, these perturbations
obey the Einstein constraint equation (a detailed derivation can
be found in~\cite{Aybat:2010sn})
\begin{equation}
  \label{eq:warpv-constraint}
  \chi' - \rtkap \phi_i' \psi_i = 0 \:,
\end{equation}
as well as the coupled second order equation
\begin{equation}
  \label{eq:warpv-pert}
  -\Psi_m''
    + \begin{pmatrix}
      2 \sigma'' &
        2 \rtkap (-\sigma' \phi_j' + \phi_j'') \\
      2 \rtkap (-\sigma' \phi_i' + \phi_i'') &
        (4\sigma'^2 - 2\sigma'') \delta_{ij} + 6\kappa^2 \phi_i' \phi_j' + V_{ij}
    \end{pmatrix} \Psi_n
    = e^{2\sigma} \Box \Psi_m \:.
\end{equation}

As in the flat case, we again perform separation of variables,
slightly abusing notation: $F(x^\mu,y)=F(y)\rho(x^\mu)$ and
$\varphi_i(x^\mu,y)=\varphi_i(y)\rho(x^\mu)$.  
Equations~\eqref{eq:warpv-constraint} and~\eqref{eq:warpv-pert} allow
us to solve for the KK tower of spin-0 modes, with
extra-dimensional profiles $F(y)$ and $\varphi_i(y)$.
Substituting these solutions in the metric and original field
variables $\Phi_i$, computing the 5D action, and then integrating
out the extra dimension yields the 4D effective action for the
mode $\rho$:
\begin{equation}
  \mcs = \int d^4x \, \mcn \left(
    - \half \partial^\mu \rho \partial_\mu \rho
    - \half m^2 \rho^2 \right)
    + \text{(surface terms)} \:,
\end{equation}
where the normalisation is
\begin{equation}
\label{eq:norm-warp-pert}
  \mcn = \int e^{2\sigma} \left( \chi^2 + \psi_i^2 \right) dy
    = \int e^{-2\sigma} \left( \frac{2}{\kappa^2} F^2 + \varphi_i^2 \right) dy \:.
\end{equation}
In deriving the effective action, we encounter three independent
surface terms of the generic form $\int S' d^4x\,dy$, where $S$ is
one of
\begin{subequations}
\begin{align}
  S_0 &= \frac{1}{3\kappa^2} e^{-4\sigma} \sigma' \:,\\
  S_1 &= \frac{1}{3\rtkap} e^{2\sigma} (e^{-4\sigma} \chi)'
    = \frac{1}{3\rtkap} e^{-2\sigma} (\chi' - 4\sigma' \chi) \:,\\
  S_2 &= \frac{-1}{6} e^{28\sigma} \chi (e^{-28\sigma} \chi)'
    - \half e^{-2\sigma} \psi_i (e^{2\sigma} \psi_i)'
    = \frac{-1}{6} \chi (\chi' - 28\sigma' \chi)
      - \half \psi_i (\psi_i' + 2\sigma' \psi_i) \:.
\end{align}
\end{subequations}
The subscripts here correspond to the order of perturbation.
The last two equations in terms of physical variables are
\begin{subequations}
\label{eq:warp-surface-terms}
\begin{align}
  S_1 &= \frac{1}{3\kappa^2} e^{2\sigma} (e^{-6\sigma} F)'
    = \frac{1}{3\kappa^2} e^{-4\sigma} (F' - 6\sigma' F) \:,\\
  S_2 &= \frac{-1}{3\kappa^2} e^{26\sigma} F (e^{-30\sigma} F)'
    - \half e^{-4\sigma} \varphi_i \varphi_i'
    = -e^{-4\sigma} \left( \frac{1}{3\kappa^2} F F' - \frac{10}{\kappa^2} \sigma' F^2 + \half \varphi_i \varphi_i' \right) \:.
\end{align}
\end{subequations}
$S_0$ must vanish on the $y$ boundary for a background
configuration to be physical.  When looking for physical modes of
perturbation, the solutions $F(y)$ and $\varphi_i(y)$ must be such
that $\mcn$ is finite and $S_{1,2}$ vanish on the boundary.

Let us now look for zero modes of this system, that is, when
$\Box \Psi=0$.  For this special case the constraint
equation~\eqref{eq:warpv-constraint} can be combined with the first
row in equation~\eqref{eq:warpv-pert} to solve for $\chi$:
\begin{equation}
  \label{eq:chi-v}
  \chi = \frac{\kappa}{\rt \sigma''}
    \left[ \phi_i' \psi_i' + (2 \sigma' \phi_i' - \phi_i'') \psi_i \right] \:.
\end{equation}
Thus the zero mode system is really just $N$ coupled, linear,
second order ODEs.  Ignoring finite normalisability and the
vanishing of the boundary terms, such a system has $2N$ linearly
independent solutions.  Two of these solutions are
\begin{subequations}
\begin{align}
  \label{eq:warpv-zm1}
  \Psi^{(1)}
    &= \begin{pmatrix} \frac{\rt}{\kappa} \sigma' \\ \phi_i' \end{pmatrix} \:,\\
  \label{eq:warpv-zm2}
  \Psi^{(2)}
    &= B_1 \Psi^{(1)}
      + \begin{pmatrix} \rtkapinv e^{-2\sigma} \\ 0 \end{pmatrix} \:,
\end{align}
\end{subequations}
where
\begin{equation}
  B_1 = \int e^{-2\sigma} dy \:.
\end{equation}
These solutions were first derived by Shaposhnikov et.\
al.~\cite{Shaposhnikov:2005hc} for the $N=1$ case (see their
equation~(3.6)), but the straightforward generalisation to all $N$
is also a solution.

For $N=1$, $\Psi^{(1)}$ and $\Psi^{(2)}$ are the two linearly
independent zero modes.  For $N>1$, we can use these known
solutions to reduce the order of the system by two.  We shall do
this explicitly for the $N=2$ case.  Begin by using
equation~\eqref{eq:chi-v} to eliminate $\chi$ in the set of
equations~\eqref{eq:warpv-pert}.  This gives two second-order
equations for $\psi_1$ and $\psi_2$.  Now write this as four
first-order equations and reduce the order by two using the method
outlined in Appendix~\ref{sec:app}.  In terms of solutions,
$f(y)$, of this reduced ODE, the final two zero modes are
\begin{equation}
  \label{eq:warpv-zm34}
  \Psi^\text{(3,4)}
    = G_1 \begin{pmatrix} \frac{\rt}{\kappa} \sigma' \\ \phi_1' \\ \phi_2' \end{pmatrix}
      + \begin{pmatrix}
          \rtkapinv e^{-2\sigma} G_2
            + \frac{\kappa}{\rt\,\sigma''} \left( 2\sigma'\phi_2'-\phi_2'' \right) \auxfun
            + \frac{\kappa}{\rt\,\sigma''} \phi_2' \auxfun' \\
          0 \\
          \auxfun
        \end{pmatrix} \:,
\end{equation}
where
\begin{subequations}
\begin{align}
  G_1 &= \int e^{-2\sigma} \, G_2 \, dy \:,\\
  G_2 &= H_1 \auxfun + \int H_2 \auxfun \, dy \:,\\
  H_1 &= 2\kappa^2 e^{2\sigma} \frac{\phi_2'}{\sigma''}
              \left[\log\left(\phi_1'e^{-\sigma}\right)\right]' \:,\\
  H_2 &= e^{2\sigma} \frac{1}{\phi_1'}
              \left( 6\kappa^2 \phi_1' \phi_2' + V_{12} \right)
        - \frac{\left(e^{-2\sigma}\phi_2'H_1\right)'}{e^{-2\sigma}\phi_2'} \:.
\end{align}
\end{subequations}
The second order ODE that the auxiliary variable $f(y)$ must solve
is
\begin{equation}
  \label{eq:auxde}
  -\auxfun'' + H_3 \auxfun' + H_4 \auxfun = 0 \:,
\end{equation}
where
\begin{subequations}
\begin{align}
    H_3
      &= \frac{d}{dy} \log\left[1+\left(\frac{\phi_2'}{\phi_1'}\right)^2\right]
      = 2\kappa^2 \frac{\phi_1'\phi_2'}{\sigma''}
           \left( \frac{\phi_2'}{\phi_1'} \right)' \:,\\
    H_4
      &= 4\sigma'^2 - 2\sigma'' + V_{22} - \frac{\phi_2'}{\phi_1'} V_{12}
           + \left( 2\sigma' - \frac{\phi_2''}{\phi_2'} \right) H_3 \:.
\end{align}
\end{subequations}
The two independent zero mode solutions, $\Psi^{(3)}$ and
$\Psi^{(4)}$, correspond to the two independent solutions for
$\auxfun$.  This system has four integration constants in total.
Those in $G_1$ and $G_2$ add to $\Psi^{(3,4)}$ a constant multiple
of $\Psi^{(1)}$ and $\Psi^{(2)}$ respectively.  The other two
constants come from the solution for $\auxfun$.

\subsubsection{Fake supergravity potential}

In the fake supergravity formalism~\cite{DeWolfe:1999cp, Freedman:2003ax}
the scalar potential is generated by a superpotential $W$:
\begin{equation}
  \label{eq:warpw-pot}
  V = \frac{1}{2} W_i^2 - 2\kappa^2 W^2 \:.
\end{equation}
The background equations are then first order:
\begin{subequations}
\begin{align}
  \sigma' &= \kappa^2 W \:,\\
  \phi_i' &= W_i \:.
\end{align}
\end{subequations}
%
As for the general $V$ case, we work with the variable
$\Psi=(\chi,\psi_i)$ for the perturbations.  The Einstein
constraint equation is
\begin{equation}
  \label{eq:warpw-constraint}
  \chi' - \rtkap W_i \psi_i = 0 \:,
\end{equation}
and the equivalent of equation~\eqref{eq:warpv-pert} factorises
to give
\begin{equation}
  \label{eq:warpw-pert}
  (\partial_y + U)(-\partial_y + U) \Psi = e^{2\sigma} \Box \Psi \:,
\end{equation}
where
\begin{equation}
  U = \begin{pmatrix}
        0 & \rtkap W_j \\
        \rtkap W_i & -2\kappa^2 \delta_{ij} W + W_{ij}
      \end{pmatrix} \:.
\end{equation}

We now look for zero mode solutions to
equations~\eqref{eq:warpw-constraint} and~\eqref{eq:warpw-pert}.
To begin with, we have the two solutions found in the general
$V$ case for all $N$, written here in terms of $W$:
\begin{subequations}
\begin{align}
  \label{eq:warpw-zm1}
  \Psi^{(1)} &= \begin{pmatrix} \rtkap W \\ W_i \end{pmatrix} \:,\\
  \label{eq:warpw-zm2}
  \Psi^{(2)} &= B_1 \Psi^{(1)}
    + \begin{pmatrix} \rtkapinv e^{-2\sigma} \\ 0 \end{pmatrix} \:.
\end{align}
\end{subequations}


We now concentrate on the $N=2$ case and perform reduction of
order on the system of equations.  Due to the factorisability of
the perturbation equation~\eqref{eq:warpw-pert}, we proceed here
in a different manner than we did in the case for general $V$.  We
begin with the third order system $(-\partial_y + U)\Psi=0$ and
use the two known solutions $\Psi^{(1,2)}$ to reduce the system
to a single first order ODE, which we solve for the third solution
$\Psi^{(3)}$.  To get the fourth solution, we take the full sixth
order system $(\partial_y+U)(-\partial_y-U)\Psi=0$ and eliminate
$\chi$ using equation~\eqref{eq:chi-v} (using $W$ instead of the
background fields).  We then use solutions $\Psi^{(1,2,3)}$ to
reduce the resulting system from order four to order one.  This
final ODE can be solved to find $\Psi^{(4)}$.  The two additional
solutions are
\begin{subequations}
\begin{align}
  \label{eq:warpw-zm3}
  \Psi^{(3)} &= C_1 \Psi^{(1)} + C_2 \Psi^{(2)}
    + \begin{pmatrix} 0 \\ \rtkapsqinv e^{-2\sigma} a_i \frac{J}{X} \end{pmatrix} \:,\\
  \label{eq:warpw-zm4}
  \Psi^{(4)} &= D_1 \Psi^{(1)} + D_2 \Psi^{(2)} + D_3 \Psi^{(3)}
    + \begin{pmatrix} \rtkapinv e^{2\sigma} \frac{W_2}{J W_1} \\ 0 \end{pmatrix} \:.
\end{align}
\end{subequations}
The auxiliary factors are
\begin{subequations}
\begin{align}
  C_1 &= \int \rt J \left( \frac{a_1 W_1 - a_2 W_2}{X} - \frac{WZ}{X^2} \right) dy \:,\\
  C_2 &= \int \left( \rtkapsqinv e^{-2\sigma} \frac{JZ}{X^2} - B_1 C_1' \right) dy \:,\\
  D_1 &= \int D_3'
    \left( -C_2 - B_1 T
      - \rtkapsqinv e^{-2\sigma} a_1 \frac{J}{W_1 X} \right) dy \:,\\
  D_2 &= \int D_3' \left( -C_1 + T \right) dy \:,\\
  D_3 &= \int \rt \kappa^2 e^{4\sigma} \frac{W'}{J^2} dy \:,\\
  T &= \frac{\rt}{\kappa} \frac{J}{W_1}
    \left[ \frac{W_2}{W'W_1} (W_1'-\kappa^2 W W_1)
      - \frac{1}{2} \frac{W_{12}}{W_1}
      + \kappa^2 a_i \frac{W}{X} \right] \:.
\end{align}
\end{subequations}
Note that everything here is ultimately defined only in terms of
$W$, its derivatives with respect to $\Phi$, and $\sigma$, all
evaluated on the background.  Once $W$ is given, everything else
can be computed in a closed form, including the four linearly
independent zero modes (for the $N=2$ case).  Also note the
identities $W'=W_1^2+W_2^2$ and $W_1'=W_{11}W_1+W_{12}W_2$.

It is not immediately obvious, but there are only three
independent integration constants in the definition of
$\Psi^{(3)}$ and four in $\Psi^{(4)}$.  These constants pull out
constant multiples of lower zero mode solutions.  In effect,
$\Psi^{(4)}$ is the most general zero mode solution.

There are two conditions that allowed us to find the general zero
mode solution in closed form for $N=2$ in the fake supergravity
case.  One, there is a constraint equation, and two, the rest of
the perturbation equations factorised.  In contrast, for the flat
case with $W$ we did not have the constraint equation, and for the
warped case with general $V$ we could not factorise.

The full zero mode solution that we derived for general $V$ in the
warped case, equation~\eqref{eq:warpv-zm34}, is equivalent to
the solution $\Psi^{(4)}$ found in this section, although they are
written in manifestly different ways.  It is straight forward to
write them in equivalent ways, allowing us to find the solutions
to equation~\eqref{eq:auxde} for $\auxfun$:
\begin{subequations}
\begin{align}
  f^{(1)} &= e^{-2\sigma} \frac{J}{W_1} D_3 \:,\\
  f^{(2)} &= e^{-2\sigma} \frac{J}{W_1} D_3 \int e^{4\sigma} \frac{W'}{J^2 D_3^2} \, dy \:.
\end{align}
\end{subequations}
(The first of these was found by inspection, the second, by
reduction of order using the first.)  Putting these solutions in
equation~\eqref{eq:warpv-zm34}, along with the relevant
substitutions for the backgrounds $\sigma$ and $\phi_i$ in terms
of $W$, yields equivalent expressions for the zero mode solutions
$\Psi^{(3,4)}$.  These can be used in place of
equations~\eqref{eq:warpw-zm3} and~\eqref{eq:warpw-zm4} if
desired.

Unfortunately we cannot use these solutions for $\auxfun$ to
intelligently deduce the correct solutions in the general $V$
case.  This is because $J$ appears in $\auxfun$, which is computed
from $W_{11}$ and $W_{22}$.  These latter functions cannot be
written in terms of $V$, its derivatives, and/or the fields
$\sigma$ and $\phi_i$.  We also remark that while one can recover
the known flat case zero mode solutions by taking $\kappa\to0$ in
the warped solutions this does not produce any new solutions for
the flat case.

\subsection{Summary of zero mode solutions}

Let us recall the main results of this section.
For $N$ scalar fields in a flat and warped extra dimension there
exist $2N$ linearly independent zero mode solutions for
perturbations around a background configuration.  These formal
solutions may or may not be physical; physicallity is obtained by
demanding the normalisation $\mcn$ is finite and the surface terms
$S_i$ vanish at the boundaries of the extra dimension.

For $N=1$ scalar field, the two zero modes are completely
determined for general $V$, for both the flat case,
equations~\eqref{eq:flatv-zm1} and~\eqref{eq:flatv-zm2-n1}, and
warped case, equations~\eqref{eq:warpv-zm1}
and~\eqref{eq:warpv-zm2}.

For $N=2$ scalars we have derived the following results:
\begin{itemize}
  \item Flat space, general $V$: one explicit solution,
    equation~\eqref{eq:flatv-zm1}.
  \item Flat space, using $W$: two explicit solutions,
    equations~\eqref{eq:flatw-zm1} and ~\eqref{eq:flatw-zm2},
    and a second order ODE for the full solution,
    equation~\eqref{eq:flatw-zm34}.
  \item Warped space, general $V$: two explicit solutions,
    equations~\eqref{eq:warpv-zm1} and~\eqref{eq:warpv-zm2},
    and a second order ODE for the full solution,
    equation~\eqref{eq:warpv-zm34}.
  \item Warped space, using $W$: four explicit solutions,
    equations~\eqref{eq:warpw-zm1}, \eqref{eq:warpw-zm2},
    \eqref{eq:warpw-zm3} and~\eqref{eq:warpw-zm4}.
\end{itemize}

For $N>2$ we can say the following for both general $V$ and a $V$
generated by a superpotential $W$.  In a flat extra dimension the
zero mode of translation always exists, given by
equation~\eqref{eq:flatv-zm1}.  Extra dimensions that are warped
admit two closed form solutions, equations~\eqref{eq:warpv-zm1}
and~\eqref{eq:warpv-zm2}.


\section{The use of zero modes}
\label{sec:use-of-zm}

By knowing the formal zero mode solutions of a system (they need
not be physically normalisable), we can deduce some important
physical properties of that system.  We shall provide some
practical remarks on using zero modes solutions to
\begin{itemize}
  \item look for linear combinations that give normalisable,
  physical zero modes;
  \item check for perturbative stability.
\end{itemize}
Our discussion here is restricted to systems whose background
configuration has definite parity, that is, $\phi_i(y)$ are either
even or odd under $y\to-y$ (different fields can have different
parities).\footnote{Note that even though the backgrounds $\phi_i$
must have parity, there is no restriction on the full field
$\Phi_i$ nor the perturbations.}.
For the warped case, $\sigma$ must always be even, due
to the Einstein equation $\sigma''=\kappa^2\phi_i'^2$.
Furthermore, since we choose $\sigma(y=0)=0$ and also demand that
at least one of the scalars have $\phi_i'(y=0)\ne0$, we have that
$\sigma$ is strictly monotonically increasing.  Then, without
negative tension branes, the extra dimension can only end when
$\sigma\to\infty$ (otherwise the junction conditions coming from
Einstein's equations cannot be satisfied at the patching points).
We then have two scenarios: either $\sigma$ diverges only as
$y\to\infty$ and we obtain an infinite extra
dimension~\cite{Randall:1999vf}, or $\sigma$ and at least one
scalar diverge at some finite $y$ value and we get a soft
wall~\cite{Gremm:2000dj, Karch:2006pv}.  Examples of both of these
types of spaces will be presented in the following sections.

The $2N$ linearly independent zero modes of a system can be
written in an infinite number of ways, as they form a basis for
the set of all solutions to the massless perturbation equation.
Regardless of the linearly independent solutions that are
obtained, one should be able to compute characteristic, physical
properties of the system in an unambiguous way.  To find these
characteristics, our idea is to construct a specific matrix of the
zero modes solutions (which will be a function of $y$) and
then compute the $y$ dependent eigenvalues of this matrix.
Looking at these eigenvalues is an extension of the idea of
looking at the determinant of the solution
matrix~\cite{AmannQuittner:1995, Berg:2006xy}.

Specifically we want to construct an $N\times N$ square matrix
whose columns are vectors of formal zero mode solutions, that is,
a single column is a vector whose $N$ entries are a particular
solution $\varphi_i$.\footnote{Recall that for the warped case the
gravitational perturbation $\chi$ could be solved for in terms of
$\psi_i$, so only the $\psi_i$ degrees of freedom, equivalently
the $\varphi_i$, are needed to construct our matrix.  Thus, the
following discussion is valid for both the flat and warped
scenarios.}
Actually, we want to construct two such
matrices: a matrix $M_E(y)$ for the even solutions~--- those that
have $\varphi_i$ the same parity as the corresponding background
field $\phi_i$~--- and $M_O(y)$ for the odd solutions, whose parity
is opposite the background field.  The initial conditions (values
of the perturbation $\varphi_i$ at $y=0$) of the solutions in the
$M_E$ ($M_O$) matrix will form a basis for an arbitrary even (odd)
mode.  For $N$ scalars there will be $N$ linearly independent even
and odd solutions, so our matrices will have $N$ columns.

As long as the above criteria are satisfied, the actual initial
conditions of the $M_{E,O}$ matrices are not important.  But for
clarity we shall describe a simple realisation.  Let the system
have $N$ background fields $\phi_i$ with parity $P_i\in\{0,1\}$
such that an even (odd) field has value 1 (0).  Then the even
solution matrix has initial conditions
\begin{equation}
  M_E(y=0) =
  \begin{pmatrix}
    P_1 & 0 & \ldots & 0 \\
    0 & P_2 & \ldots & 0 \\
    \vdots & \vdots & \ddots & \vdots \\
    0 & 0 & \ldots & P_N \\
  \end{pmatrix} \:,
  \qquad
  M_E'(y=0) =
  \begin{pmatrix}
    1-P_1 & 0 & \ldots & 0 \\
    0 & 1-P_2 & \ldots & 0 \\
    \vdots & \vdots & \ddots & \vdots \\
    0 & 0 & \ldots & 1-P_N \\
  \end{pmatrix} \:.
\end{equation}
The odd solution matrix has opposite initial conditions:
$M_O(y=0)=M_E'(y=0)$ and $M_O'(y=0)=M_E(y=0)$.  Given these initial
values, one must then compute all entries in the two matrices as
a function of $y$, up to the boundary of the extra dimension.
This can be accomplished by using the closed form expressions for
the zero modes in the previous section, or by directly integrating
the coupled ODEs describing the perturbations.  In the former case
the integration constants in the closed form solutions must be
chosen to achieve the correct initial conditions.  In the latter
case the initial conditions in $M_{E,O}$ can be used directly as
initial conditions in, for example, a numerical ODE solver.

The full $M_{E,O}(y)$ matrices form a basis of even and odd
solutions because their initial conditions form a basis of all
possible initial conditions.  The set of all zero mode solutions
is generated by the matrix products $M_E\cdot\alpha$ and
$M_O\cdot\alpha$, where $\alpha = (\alpha_1,\ldots,\alpha_N)$,
$\alpha_i\in\mathbb{R}$ is a vector of constant coefficients.

We can now compute the eigenvalues of $M_{E,O}$.  They will be
functions of $y$ and there will be $N$ of them; let us denote them
by $\lambda^{E,O}_i(y)$.  These eigenvalue functions give us a lot
of information about the stability of our background
configuration.  Assuming that the condition for normalisability is
that a perturbation must vanish at the boundaries of the extra
dimension, at $y_*$, we make the following two conjectures:
\begin{enumerate}
  \item {\it For each $\lambda^E_i\to0$ as $y\to y_*$ there exists
  a corresponding, normalisable, even zero mode given by
  $M_E\cdot\alpha_i$, where $\alpha_i$ is the eigenvector for
  $\lambda^E_i(y_*)$.  The correspondence is one-to-one: the
  existence of a normalisable mode implies the vanishing of one of
  the eigenvalues at $y_*$.  An equivalent statement is true for
  the odd sector.}
  \item {\it The number of negative mass modes in the full
  spectrum of perturbations equals the number of times the
  eigenvalues $\lambda^{E,O}_i$ pass through zero in the domain
  $0\le y<y_*$.}
\end{enumerate}

We shall sketch the proof for the first conjecture.
Let $M$ be either $M_E$ or $M_O$.  A normalisable zero mode exists
if we can find a linear combination of formal mode solutions that
vanishes as $y\to y_*$.\footnote{Since we are dealing with a
Schr\"odinger-like equation, solutions at infinity either
oscillate or behave exponentially.  If a solution asymptotes to
zero then it must decay exponentially, implying square
integrability.}
That is, we want to find a constant,
non-trivial vector $\alpha$ such that $M\cdot\alpha\to0$ for
$y\to y_*$.  This is simply an eigenvalue equation with a zero
eigenvalue, and with eigenvector $\alpha$.  Thus, we want to find
the $N$ eigenvalues of the solution matrix $M$, called
$\lambda_i(y)$, and we want at least one of these eigenvalue
functions to tend to zero at the boundary of $y$.  If there exists
such a $\lambda_i$, then the corresponding eigenvector evaluated
at $y_*$ gives the coefficients needed to construct a normalisable
zero mode.  Conversely, if a normalisable zero mode is known to
exist, then one can find the vector $\alpha$ such that
$M\cdot\alpha=0$ at $y=y_*$, and so $M$ has an eigenvalue function
which vanishes at the boundary.

For the second conjecture we do not provide a proof.  It is based
on a closely related theorem given by Amann and
Quittner~\cite{AmannQuittner:1995}.  They work with a system of
coupled radial Schr\"odinger equations and the wavefunction values
must all vanish at the origin; effectively they are looking only
for solutions where all wavefunctions are odd.  Their proof should
be adaptable to our second conjecture stated above, including correct
handling of the weight function $e^{2\sigma}$ in the warped case
in equation~\eqref{eq:warpv-pert}.  We do not attempt to construct
the proof here, and the analysis in the following sections is
largely independent of it.  Our interest in presenting the second conjecture
is to show that the eigenvalues $\lambda_i^{E,O}$ contain more
information than just whether or not a normalisable zero mode
exists.  For an application of Amann and Quittner's theorem to a
system with 16 components see~\cite{Garaud:2010ng}.

To summarise, we claim that the eigenvalue functions
$\lambda_i^{E,O}(y)$ of the general solution matrices $M_{E,O}(y)$
give all the information about the perturbative stability of a
specific background configuration, for both flat and warped extra
dimensions.  They tell the number of unstable modes, if any, and
whether or not the configuration is critically stable, that is,
it admits a normalisable zero mode.  For phenomenological
reasons, one generally tries to construct models that are free of
zero modes.

In the following sections we shall apply our first conjecture to
some example models to show the existence, or lack thereof, of
zero modes.

Before moving on to the examples, let us discuss the massless
translation mode associated with a background.  For the flat case,
there is always a formal solution corresponding to translations of
the background, equation~\eqref{eq:flatv-zm1}.  Assume this
solution is normalisable.  We would like to know what happens to
this mode when one adds gravity to the system.  For $N=1$ scalar,
Shaposhnikov et al~\cite{Shaposhnikov:2005hc} have shown that when
gravity is turned on this mode is no longer massless and becomes
instead a wide resonance.  

Now consider the existence of a translation mode for general $N$
with gravity.  The solution $\Psi^{(1)}$ given by
equation~\eqref{eq:warpv-zm1} looks like it has the right form
for such a mode, as the $\psi_i$ components are exactly $\phi_i'$.
Even though the actual perturbations are
$\varphi_i=e^{2\sigma}\psi_i=e^{2\sigma}\phi_i'$, this solution
still has the correct initial conditions for it to be a valid
translation mode: $\varphi_i(0)=\phi_i'(0)$ and
$\varphi_i'(0)=\phi_i''(0)$ since $\sigma(0)=\sigma'(0)=0$.
These initial conditions are enough to uniquely specify the mode
solution, so $\Psi^{(1)}$ \emph{is} the solution with the initial
conditions of a translation mode.  But this mode is
non-normalisable, since $\sigma\to\infty$ as
$y\to y_*$.\footnote{Unless there exists $\phi_i$ such that
$e^{2\sigma}\phi_i'\to0$ as $y\to y_*$, with
$\sigma''=\kappa^2\phi_i'^2$.}
We therefore conclude that for general $N$ with a warped extra
dimension the usual translation mode is rendered non-normalisable
by the introduction of gravity.

From a phenomenological point of view the absence of a translation
mode is good news, since massless spin-0 particles are not seen in
nature.  But, for $N>1$, it may be that there are additional zero
modes in the spectrum that survive when gravity is turned on.  For
example, there may exist a zero mode that both translates and
dilates the background, and so has different initial conditions
to $\Psi^{(1)}$ and is normalisable.  We show in the following
sections that such modes do in general exist, and the addition of
gravity does not in general remove all zero modes from the spin-0
particle spectrum.


\section{Domain-walls in an infinite extra dimension}
\label{sec:dw}

Our first example model has an infinite extra dimension with $N=2$
scalars in a kink-lump configuration.  This type of set-up was
used in~\cite{Davies:2007xr} in an attempt to realise the standard
model confined to a domain-wall.  We shall present two
incarnations of this model.  The first is effectively that
presented in~\cite{Davies:2007xr} and is described by a
straightforward quartic potential; we shall call it the $V$-model.
As will be shown, in the flat case this $V$-model has a single
zero mode, whereas in the warped case the zero mode is absent.
The second incarnation is based on the fake supergravity approach
and is described by a superpotential $W$; we call this model the
$W$-model.  We shall see that although the background solutions are
qualitatively the same as those in the $V$-model (and in fact can
be made exactly the same for certain choices of parameters) the
$W$-model posses two zero modes in the flat case, one of which
survives when gravity is turned on.

\subsection{Kink-lump model in flat space}

The Lagrangian of the flat-space $V$-model is given by
equation~\eqref{eq:lag} with $N=2$ scalars, $\Phi_{1,2}$, and
potential
\begin{equation}
  \label{eq:flat-v}
  V =
    - 2 l v^2 \, \Phi_1^2
    - \half \mu_{\chi}^2 \, \Phi_2^2
    + \half c \, \Phi_1^2 \Phi_2^2
    + l \, \Phi_1^4
    + \frac{\lambda}{4} \, \Phi_2^4
    + l v^4 \:.
\end{equation}
It is not our aim here to perform a complete analysis of this
model.  Instead, we shall use it to give a constructive example of
how one looks for normalisable zero modes, and show the effect of
adding gravity.

We first restrict ourselves to the region of parameter space where
all five parameters are positive, $c v^2 - \mu_{\chi}^2 > 0$ and
$4 l \lambda v^4 - \mu_{\chi}^4 > 0$.  This ensures that the
kink-lump configuration is stable (has no negative
modes)~\cite{Davies:masters}.  Next, we impose the relation
$2(\lambda-c)\mu_{\chi}^2 = (2 c \lambda - 4 l \lambda - c^2) v^2$
which allows us to obtain analytic solutions for the
background~\cite{Davies:2007xr}:
\begin{equation}
  \label{eq:flat-su5-bg}
  \phi_1 = v \tanh(k y) \:,
  \qquad
  \phi_2 = A \cosh^{-1}(k y) \:,
\end{equation}
where $k^2 = c v^2 - \mu_{\chi}^2$ and
$A^2 = (2 \mu_{\chi}^2 - c v^2)/\lambda$.  These are solutions of
equation~\eqref{eq:flatv-bg}.  The background configuration is
straightforward with $\phi_1$ the kink and $\phi_2$ the lump;
we do not plot these functions for the flat case, but see
Figure~\ref{fig:warp-su5-bg} for qualitatively similar plots of
$\phi_{1,2}$ in the warped case.  Note that in all our plots we
show only the $y\ge0$ half of the fields.  The other halves are
found by relevant parity transformations.

\begin{figure}
  \centering
  \includegraphics[width=.49\textwidth]{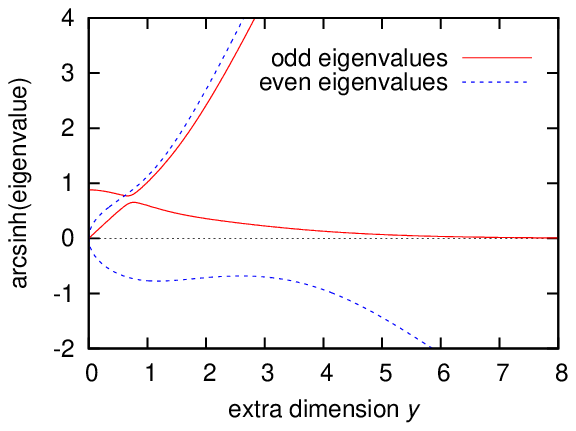}
  \hfill
  \includegraphics[width=.49\textwidth]{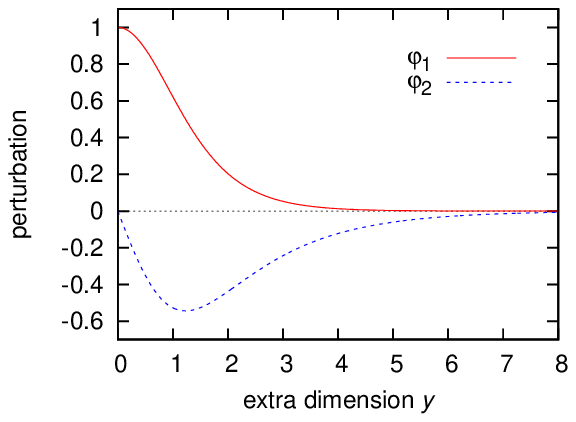}
  \caption{
    Eigenvalues of the zero-mode solution matrices $M_{E,O}(y)$
    (left) and the normalisable zero mode of translation (right)
    for the flat-space kink-lump $V$-model.  The single odd
    eigenvalue that asymptotes to zero at large $y$ signals the
    existence of the translation zero mode.  Since the other three
    eigenvalues diverge, there are no other normalisable zero
    modes.
  }
  \label{fig:flat-su5-evals-zm}
\end{figure}

Using this background, we now solve equation~\eqref{eq:flatv-pert}
for the four independent zero modes, two which are even and two
which are odd, and construct the zero-mode solution matrices
$M_{E,O}(y)$.  Recall that even and odd are relative to the
background configuration, so, for example, the even set of zero
modes in $M_E(y)$ have $\varphi_1$ odd and $\varphi_2$ even.
Parameters we choose are
$l=0.7$, $v=1.0$, $c=1.5$ and $\lambda=0.4$,
with derived parameter
$\mu_{\chi}^2=0.98636$.
This choice gives typical looking solutions which exhibit the
behaviour we are interested in.  It is not a fine-tuned
choice and small variations in the parameters give qualitatively
similar results.  Having computed the two $2\times 2$ matrices
$M_{E,O}(y)$ we then compute their two eigenvalues (four
eigenvalues in total), which are plotted as a function of $y$ on
the left in Figure~\ref{fig:flat-su5-evals-zm}.  In order to
accommodate the large range of the eigenvalues $\lambda_i^{E,O}$ we
implement a quasi-log scale by plotting
$\arcsinh[\lambda_i^{E,O}(y)]$.

This plot gives a quantitative summary of the stability behaviour
of the background configuration of the $V$-model.  According to
conjecture one, the existence of an odd eigenvalue $\lambda^O$
with $\lambda^O\to0$ as $y\to\infty$, as is apparent in the plot,
implies the existence of a normalisable zero mode.  The
eigenvector corresponding to $\lambda^O$ evaluated at large $y$
gives the linear combination of formal zero mode
solutions which yield the normalisable zero mode.  The resulting
initial conditions for this normalisable mode are
$(\varphi_1(0),\varphi_2(0))=(1,0)$ and
$(\varphi_1'(0),\varphi_2'(0))=(0,-0.77912)$, and the mode is
plotted on the right in Figure~\ref{fig:flat-su5-evals-zm}.  This
mode is exactly the translation zero mode given by
equation~\eqref{eq:flatv-zm1}, as expected.  The fact that the
other three eigenvalues diverge at large $y$ implies, by conjecture
one, that there are no more normalisable zero modes for this
particular background configuration.  By conjecture two, since none
of the eigenvalues cross zero there are no negative mass modes,
which is again as expected due to our choice of parameters.

Let us now consider the flat $W$-model, which admits similar
kink-lump configurations as the $V$-model, but has different
behaviour when it comes to the zero modes.  The $W$-model has a
potential described by equation~\eqref{eq:flatw-pot} with
superpotential given by
\begin{equation}
  \label{eq:w-kink-lump}
  W = a \, \Phi_1 - b \, \Phi_1^3 - c \, \Phi_1 \Phi_2^2 \:.
\end{equation}
This model \emph{always} has the analytic kink-lump background
solution given by equation~\eqref{eq:flat-su5-bg}, with the
constants in the solution given in terms of the parameters in $W$
as $v^2 = a/3b$, $k^2 = 4ac^2/3b$ and $A^2 = a/c - 2a/3b$.  For
the following analysis we choose parameters to give the same $v$,
$k$ and $A$ as in the $V$-model, namely $a=1.14017$, $b=0.38006$
and $c=0.35834$.  The outcome of our analysis is not so dependent
on parameter choice since we simply want to demonstrating the
existence of zero modes.

\begin{figure}
  \centering
  \includegraphics[width=.49\textwidth]{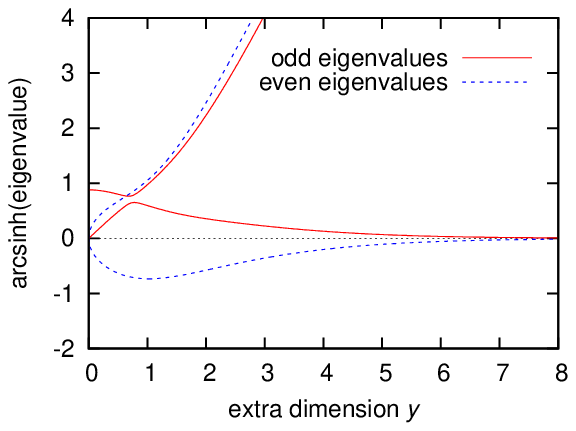}
  \hfill
  \includegraphics[width=.49\textwidth]{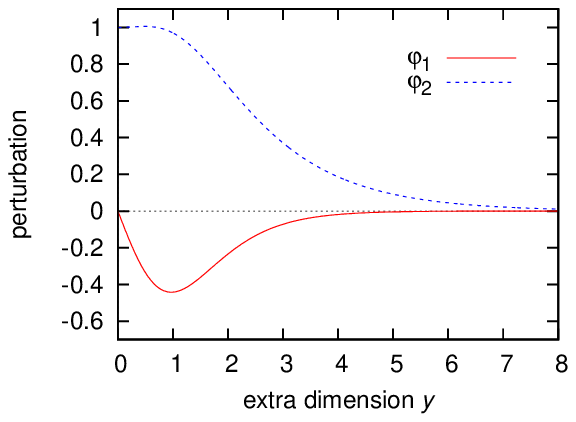}
  \caption{Eigenvalues of the solution matrices (left) and the
    normalisable even zero mode (right) for the flat-space
    $W$-model.  There are two normalisable zero modes for this
    model, the other one is the odd translation mode and is
    exactly the same as in the $V$-model.
  }
  \label{fig:flat-wsu5-evals-zm}
\end{figure}

As before, we compute the solution matrices and their eigenvalues,
the latter of which are shown on the left in
Figure~\ref{fig:flat-wsu5-evals-zm}.  From this plot we see that
the $W$-model has two normalisable zero modes, one odd and one
even, and no negative modes.  The normalisable odd mode is the
translation mode, and is the same as in the $V$-model.  The
normalisable even mode has initial conditions
$(\varphi_1(0),\varphi_2(0))=(0,1)$ and
$(\varphi_1'(0),\varphi_2'(0))=(-0.77912,0)$, and is shown on the
right in Figure~\ref{fig:flat-wsu5-evals-zm}.  Physically, this
even mode expands both $y$ and $\phi_2$, or shrinks both.
Although any linear combination of the odd and even normalisable
zero modes is again a normalisable zero mode, what is important is
that there are a total of two massless physical degrees of
freedom.  The choice between an odd and even basis, or some other
basis without parity, will depend on the physical problem at hand.

Note that in~\cite{Aybat:2010sn} it is shown that for models
generated by a superpotential $W$, exciting the zero modes
corresponds to changing the integration constants in the first
order equations of motion for the background, $\phi_i'=W_i$.
In our example here we have two fields, two integration constants,
and so two zero modes.  The odd mode changes the value $\phi_1(0)$
and the even mode changes $\phi_2(0)$.  These modes are massless
because changes in the initial values $\phi_i(0)$ do not change
the energy density of the configuration.

In summary, even though the $V$-model and $W$-model have the
same background configuration, the former admits only one
normalisable zero mode, while the latter admits two.  Since the
$W$-model contains some extra symmetries owing to its
supersymmetric nature, it has the extra zero mode.  These results
are instructive, although not particularly profound in their own
right.  The reason we have constructed these two models is so we
can compare, qualitatively, what happens when gravity is turned on
and the extra dimension is warped.

\subsection{Kink-lump model in warped space}

We now look at the existence of zero modes for the kink-lump model
when gravity is turned on and the infinite extra dimension is
warped, as per a smoothed-out version~\cite{DeWolfe:1999cp,Csaki:2000fc}
of Randall-Sundrum~\cite{Randall:1999vf}.  We shall analyse both
the $V$- and $W$-models.

\begin{figure}
  \centering
  \includegraphics[width=.6\textwidth]{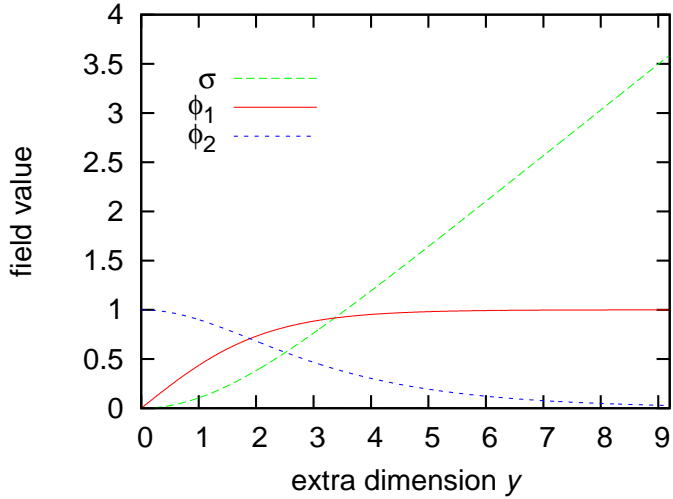}
  \caption{Background configuration for the kink-lump model with
    a warped extra dimension.
  }
  \label{fig:warp-su5-bg}
\end{figure}

The action is given by equation~\eqref{eq:warp-act} with metric
ansatz~\eqref{eq:warp-metric}.  The scalar potential is
\begin{equation}
  V_\text{warp} = V + \Lambda \:,
\end{equation}
where $V$ is the flat space potential, equation~\eqref{eq:flat-v},
and $\Lambda$ is the bulk cosmological constant required to
fine-tune flat 4D slices in the warped space.  For stable
solutions, constraints on the parameters in the potential are the
same as for the flat space case.  We can again obtain analytic
background solutions, but the relations are now different, being
\begin{subequations}
\begin{align}
  \lambda &= 2 c - 4 l \:,\\
  \mu_{\chi}^2 &= \frac{l v^2}{1 + 2 \kappa^2 v^2}
    + \frac{\lambda v^2 (3 + 8 \kappa^2 v^2)}{4 + 8 \kappa^2 v^2} \:,\\
  \Lambda &= -2 \kappa^2 v^4 k^2 \:,
\end{align}
\end{subequations}
where $k^2 = (c v^2 - \mu_{\chi}^2)/(1 + 4 \kappa^2 v^2)$.  Recall
that $\kappa$ is the 5D Newton's constant.  With these relations
the background configuration is
\begin{equation}
  \label{eq:warp-su5-bg}
  \sigma = \kappa^2 v^2 \log\left[\cosh(k y)\right] \:,
  \qquad
  \phi_1 = v \tanh(k y) \:,
  \qquad
  \phi_2 = v \cosh^{-1}(k y) \:.
\end{equation}
A representative choice of parameters is
$\kappa=1.0$, $l = 0.7$, $v = 1.0$ and $c = 1.5$,
with derived parameters
$\lambda=0.2$, $\mu_{\chi}^2=0.41667$.
Plots of the background fields are given in Figure~\ref{fig:warp-su5-bg}.
In the gravity-free limit $\kappa\to0$, and we have $\sigma\to0$
while $\phi_{1,2}$ retain their kink-lump structure, so we obtain
similar solutions as in the flat space case.  But we must make it
clear that the point in parameter space we have chosen for the
warped case is not the same (but it is close to) the point in
parameter space that we analysed in the flat case.  Nevertheless,
since we are interested only in qualitative features of the
set-ups, we can still make a fair comparison between the flat and
warped configurations.

Given this warped background we can proceed to compute the zero
mode solution matrices $M_{E,O}(y)$.  The metric perturbation
$\chi$ (and its counterpart $F$) can be written in terms of the
$\psi_i$ (counterparts $\varphi_i$), so we only need the latter to
construct the solution matrices.  Now, if we use $\psi_i$ to
construct the matrices and look for eigenvalues that tend to zero
for large $y$, we will obtain modes that are normalisable with
respect to the integral $\int \psi^2 dy$.  But what we really want
are modes that are normalisable as per
equation~\eqref{eq:norm-warp-pert}.  So in fact we should
construct solution matrices from $\varphi_i$, which, by
equation~\eqref{eq:varphi-psi-relation}, is simply
$e^{2\sigma} M_{E,O}(y)$, where $M_{E,O}$ here is constructed from
$\psi_i$.

\begin{figure}
  \centering
  \includegraphics[width=.6\textwidth]{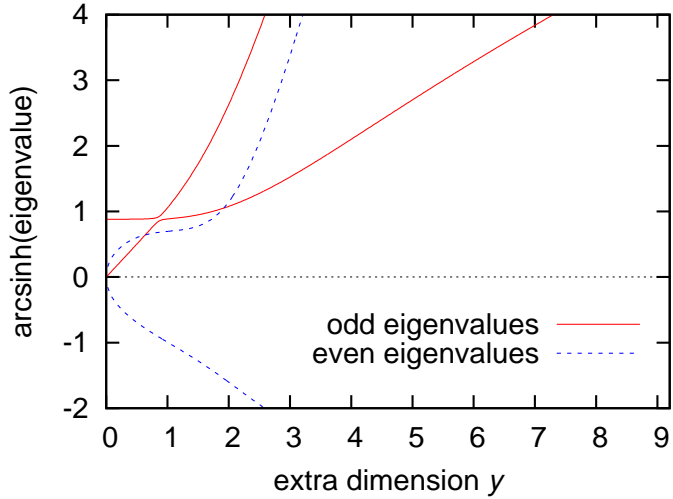}
  \caption{Eigenvalues for odd and even zero modes for the
    $V$-model with a warped extra dimension.  All eigenvalues
    diverge for large $y$ so there are no normalisable zero modes.
    Since the eigenvalues do not pass through zero, so there are
    also no negative modes.
  }
  \label{fig:warp-su5-evals}
\end{figure}

Figure~\ref{fig:warp-su5-evals} shows the eigenvalues of the
solutions for the warped $V$-model.  All of the eigenvalues
diverge for large $y$ so there are no zero modes in the physical
spectrum.  In particular, the translation zero mode does not
survive in the presence of gravity, in concordance with the
results of~\cite{Shaposhnikov:2005hc}.  Similarly, since none of
the eigenvalues pass through zero, there are no negative modes in
the spectrum either.  This is also as expected since we restricted
our parameters so the configuration would be stable.  In summary,
the $V$-model in warped space has no normalisable negative modes
and no normalisable zero modes.

Now consider the $W$-model in warped space.  We use the same
superpotential as given by equation~\eqref{eq:w-kink-lump}, but
now the derived potential $V$ is modified, as per
equation~\eqref{eq:warpw-pot}.  Such a potential is again, as in
the flat space case, qualitatively similar to the $V$-model.  In
fact, for the choice of parameters $b=c$, we can get analytic
solutions of exactly the same form as the warped $V$-model,
equation~\eqref{eq:warp-su5-bg} (such a model is used
in~\cite{Bazeia:2004dh}).  The parameters of this solution in
terms of the parameters in $W$ are $v^2=a/3b$ and $k^2=4ab/3$.  To
obtain a background with exactly the same form as the one we used
in the analysis of the $V$-model we choose $a=0.69821$ and
$b=c=0.23274$; the background is show in
Figure~\ref{fig:warp-su5-bg}.  The conclusions that we shall draw
regarding zero modes are generically the same for a large
parameter range, but we make this choice so we can compare with
the $V$-model.

\begin{figure}
  \centering
  \includegraphics[width=.49\textwidth]{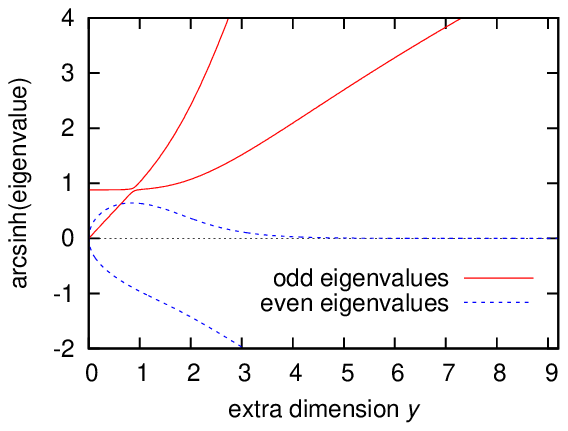}
  \hfill
  \includegraphics[width=.49\textwidth]{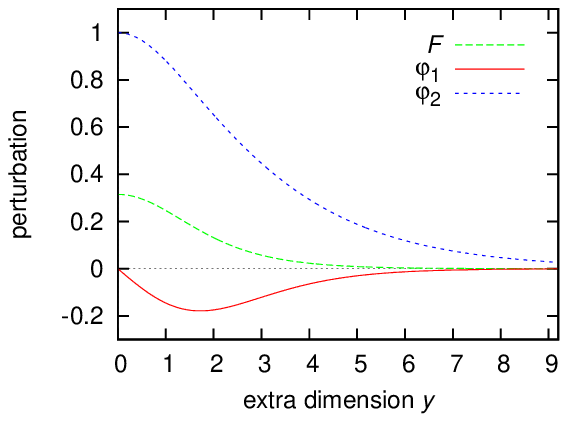}
  \caption{
    Eigenvalues of the solution matrix (left) and the single
    normalisable zero mode (right) for the warped-space $W$-model.
    Compare with the flat-space $W$-model,
    Figure~\ref{fig:flat-wsu5-evals-zm}, which has two zero modes,
    and the warped-space $V$-model,
    Figure~\ref{fig:warp-su5-evals}, which has no zero modes.
  }
  \label{fig:warp-wsu5-evals-zm}
\end{figure}

The eigenvalues of the solution matrix for the physical
perturbations $\varphi_i$ for the warped $W$-model are shown on
the left in Figure~\ref{fig:warp-wsu5-evals-zm}.  As can be seen,
there is a surviving even zero mode.  The initial conditions for
this mode are $(\varphi_1(0),\varphi_2(0))=(0,1)$ and
$(\varphi_1'(0),\varphi_2'(0))=(-0.172148,0)$ and the mode is
plotted on the right in Figure~\ref{fig:warp-wsu5-evals-zm}.  It
is normalisable, as per equation~\eqref{eq:norm-warp-pert}, and
the surface terms, equation~\eqref{eq:warp-surface-terms}, vanish
at large $y$.  The mode is therefore present in the physical
spectrum.  It has a qualitatively similar form to the even zero
mode in the flat $W$-model.  Finally, no eigenvalue crosses zero,
so there are no negative modes in the spectrum, a result which is
already known for general $W$~\cite{Aybat:2010sn}.

Even though the potentials of the four models we have looked at
(flat and warped, $V$- and $W$-models) are qualitatively very
similar and admit the same background configurations, they show
very different behaviour when it comes to having zero modes in the
physical spectrum.  Our general conclusion is that adding gravity
in the form of a warped extra dimension will remove the
translation zero mode from the spectrum, but will not necessarily
remove other zero modes.  In this section we have also explicitly
shown how the eigenvalues of the solution matrices $M_{E,O}(y)$
allow one to easily find normalisable zero modes.  Furthermore,
our analysis of the warped $V$-model shows, at least for some
values of the parameters, that it contains no zero modes.  It is
therefore phenomenologically acceptable to use this type of set-up
for constructing realistic models, as is done
in~\cite{Davies:2007xr}.


\section{Domain-wall soft-wall models}
\label{sec:sw}

In this section we briefly analyse another example model, one with
a compact extra dimension and $N=2$ scalars.  The potential is
generated by a superpotential with gravity and we show that a zero
mode again survives in this compact set-up.  The model was first
presented in~\cite{Aybat:2010sn} as a realisation of a domain-wall
soft-wall model, where the extra dimension is dynamically
compactified by the formation of curvature singularities.

\begin{figure}
  \centering
  \includegraphics[width=.6\textwidth]{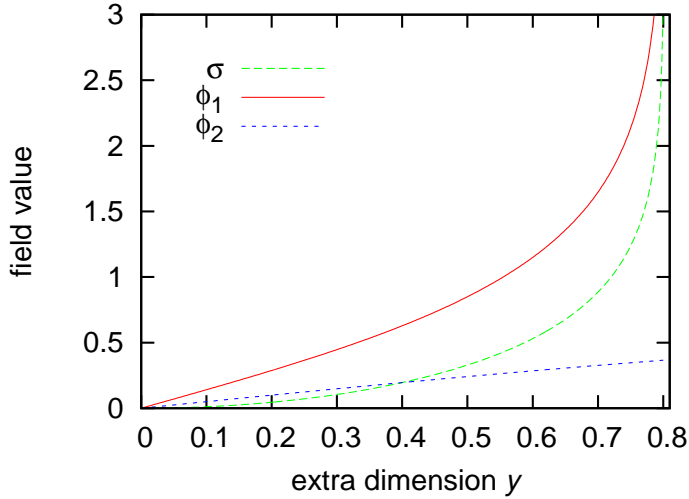}
  \caption{Domain-wall soft-wall background configuration.
    $\phi_1$ acts like a dilaton, while $\phi_2$ takes on a (very
    wide) domain-wall solution.  The extra dimension ends at a
    physical curvature singularity at $y_*=0.80143$.
  }
  \label{fig:warp-wsinh-bg}
\end{figure}

The superpotential is
\begin{equation}
  W = \alpha \sinh(\nu \, \Phi_1) + (a \, \Phi_2 - b \, \Phi_2^3) \:.
\end{equation}
For background configurations of definite parity with both scalars
odd there is a unique solution to the first order equations of
motion:
\begin{subequations}
\begin{align}
  \sigma &= \frac{-\kappa^2}{\nu^2} \log\left[\cos\left(\alpha \nu^2 y\right)\right]
    + \frac{\kappa^2 a}{18 b} \left\{
      1
      + 4 \log\left[\cosh\left(\sqrt{3 a b} \, y\right)\right]
      - \cosh^{-2}\left(\sqrt{3 a b} \, y\right)
    \right\} \:,\\
  \phi_1 &= \frac{2}{\nu} \arctanh\left[\tan\left(\frac{\alpha \nu^2 y}{2}\right)\right] \:,\\
  \phi_2 &= \sqrt{\frac{a}{3 b}} \tanh\left(\sqrt{3 a b} \, y\right) \:.
\end{align}
\end{subequations}
The edge of extra dimension is fixed at $y_* = \pi/2\alpha\nu^2$.
We choose parameters $\alpha = 1.0$, $\nu = 1.4$, $a = 0.5$ and
$b = 0.3$ and plot the background in
Figure~\ref{fig:warp-wsinh-bg}.
In~\cite{Aybat:2010sn} it was shown that enforcing odd parity on
the \emph{fields} $\Phi_{1,2}$ themselves, as opposed to just the
background configuration, ensures that there are no normalisable
zero modes in the spectrum.  We shall now show that relaxing the
parity condition leads to the appearance of a zero mode that
de-stabilises the background.

\begin{figure}
  \centering
  \includegraphics[width=.49\textwidth]{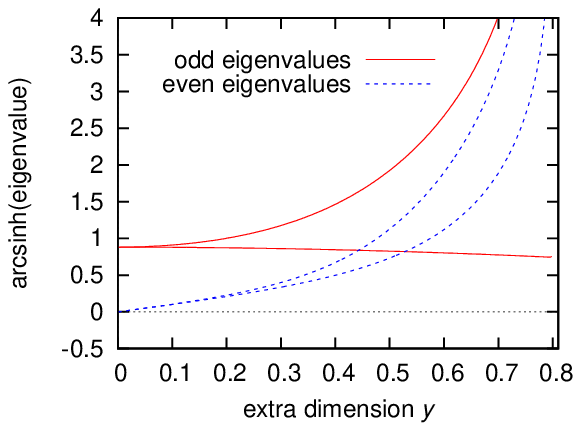}
  \hfill
  \includegraphics[width=.49\textwidth]{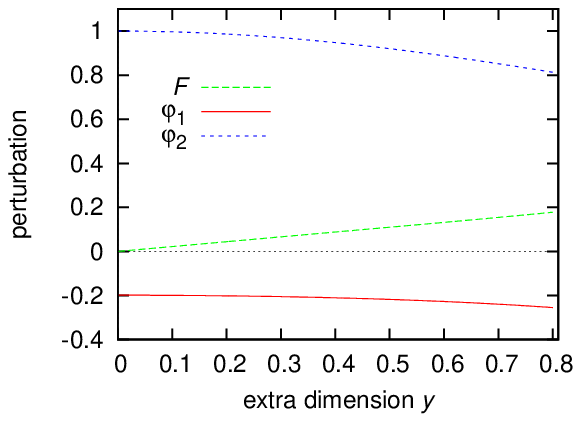}
  \caption{
    Eigenvalues of the solution matrices (left) and the
    normalisable odd zero mode (right) for the domain-wall
    soft-wall model.
  }
  \label{fig:warp-wsinh-evals-zm}
\end{figure}

For the background configuration and choice of parameters given
above we compute the solution matrices $M_{E,O}(y)$ for the
physical perturbations $\varphi_i$.  The eigenvalues of these
two matrices are show on the left in
Figure~\ref{fig:warp-wsinh-evals-zm}.  Three of the eigenvalues
diverge as $y\to y_*$, but the other one remains finite.  Even
though our first conjecture states that we must look for
eigenvalues that tend to zero to find normalisable modes, we find
that this finite eigenvalue does actually correspond to a properly
normalisable mode.  Since our space is finite, if $\varphi_i$ and
$F$ remain finite throughout the extra dimension they will be
physically allowed perturbations (a small multiple of them will be
small compared with $\phi_i$ and $\sigma$).  The initial
conditions for the normalisable mode corresponding to the finite
odd eigenvalue are $(\varphi_1(0),\varphi_2(0))=(-0.19847,1)$ and
$(\varphi_1'(0),\varphi_2'(0))=(0,0)$.  This mode is shown on the
right in Figure~\ref{fig:warp-wsinh-evals-zm}.  It is normalisable
as per equation~\eqref{eq:norm-warp-pert}, and the relevant
surface terms in the effective 4D action vanish because
$e^{-4\sigma}\to0$ at the boundaries of the extra dimension.  This
zero mode has opposite parity to the background, so if one does
not enforce parity on the fields themselves this mode will exist
in the 4D spectrum and the background configuration will not be
stable.

As we have shown, the eigenvalues of the solution matrices are
also useful for finding zero modes when the extra dimension is
compact and the perturbations are allowed to be finite over all
$y$.  The domain-wall soft-wall model in~\cite{Aybat:2010sn}
contains a normalisable zero mode which must be removed by
enforcing parity on the fields themselves (the model-building
philosophy of these set-ups forbids adding fundamental branes to
restrict the boundary conditions of the KK modes, thereby
eliminating any zero modes).  Alternatively, we suggest that it
may be possible to construct a domain-wall soft-wall model using
a normal potential $V$ that does not have the extra symmetries
inherent in the superpotential approach, and hence does not have a
surviving zero mode.


\section{Conclusions}
\label{sec:concl}

In this paper we looked at scalar perturbations around a
background configuration, and were concerned with finding zero
modes.  Both flat and warped extra dimensions were studied, with
both a general scalar potential $V$, and a potential generated by
a superpotential $W$.  The results can be split into three main
parts: analytic zero mode solutions, use of these solutions, and
examples.
In Section~\ref{sec:zm} we presented analytic
expressions for formal zero mode solutions to the perturbation
equations, with particular emphasis on the $N=2$ scalar case.
For $N=2$ scalars in warped space with a potential generated by a
superpotential we found analytic closed-form expressions for the
four, linearly independent zero modes solutions.
Section~\ref{sec:use-of-zm} discusses the use of formal zero
modes, and here we made two conjectures which use the
eigenvalues of the solution matrices: one states how to find
normalisable zero modes, the other tells how many normalisable
negative modes there are in the spectrum.
In Sections~\ref{sec:dw} and~\ref{sec:sw} we looked at some
specific example models to demonstrate the workings of these
conjectures, and to show that normalisable zero modes can survive
when the extra dimension is warped.

Our general conclusions regarding extra-dimensional model building
are the following.  If one uses a general potential $V$ then in
flat space there will exist the translation zero mode, which is
removed from the 4D KK spectrum when the extra dimension is
warped (for $N=1$ scalar this conclusion was made
in~\cite{Shaposhnikov:2005hc}).  But it is not always the case
that all spin-0 zero modes are removed by the inclusion of
gravity.  We have shown that models which have $N=2$ scalars and
generate the potential from a superpotential generally admit an
extra zero mode which survives in the presence of
gravity.\footnote{For related results on the survival of light
(but not exactly massless) spin-0 states,
see~\cite{Elander:2009pk, Elander:2010wd}.}
Such superpotential models are widely studied in the literature
for the reason that they give first-order equations of motion.
But without some extra input, like a fundamental brane or forced
parity, these superpotential models will be phenomenologically
unacceptable due to the presence of massless spin-0 degrees of
freedom, something which we have not observed in nature.

Finally, the general solutions we have found for the fake
supergravity scenario with $N=2$ scalars rely only on the
metric ansatz~\eqref{eq:warp-metric} and should have wider
applicability than to the domain-wall models that we emphasise
here.  For example, the inclusion of fundamental brane terms would
change only the boundary conditions; the bulk solutions we have
found will remain the same.  Extending the zero mode solutions to
more than one extra dimension may also be possible, following the
analysis of~\cite{Underwood:2010pm}.


\begin{acknowledgments}
We would like to thank M.~Postma for comments on a draft of this
manuscript.
This research was supported by the Netherlands Foundation for
Fundamental Research of Matter (FOM) and the Netherlands
Organisation for Scientific Research (NWO).
\end{acknowledgments}


\appendix

\section{Reduction of order of a set of linear homogeneous
ordinary differential equations}
\label{sec:app}

Any set of linear homogeneous ODEs which is $n^\text{th}$ order
can be easily recast as a set of $n$ first order differential
equations.  We assume this has been done and that the resulting
$n$ dependent variables $f_i(y)$, where $i=1\ldots n$ and $y$ is
the independent variable, satisfy the equations
\begin{equation}
  \label{eq:f-ode}
  f_i'(y) = A_{ij}(y) f_j(y) \:.
\end{equation}
There is an implicit sum over $j$.  This equation has $n$
independent solutions.  If we know $m$ of these solutions then we
can perform reduction of order on~\eqref{eq:f-ode} to obtain a set
of $n-m$ coupled first order ODEs.  We outline this procedure,
which follows closely that given
in~\cite{CoddingtonLevinson:1955}.

Write equation~\eqref{eq:f-ode} as a matrix equation,
$f'(y) = A(y) \cdot f(y)$, where $f(y)$ is a vector of length $n$
made of the dependent variables.  Let $F_s$ be the $n\times n$
solution matrix of this ODE, such that each column of $F_s$ is an
independent solution vector $f^{(s)}(y)$.  It must be that
$\det F_s \ne 0$ for all $y$.  Then $F_s$ is known as the
fundamental matrix of the set of ODEs specified by $A$, since
$F_s$ determines $A$ uniquely by $A=F_s' F_s^{-1}$ (the converse
is not true since $F_s \cdot (\text{const matrix})$ is also a
fundamental matrix of $A$).

Reduction of order then proceeds as follows.  Assume we know $m$
columns of $F_s$ (that is, $m$ linearly independent solutions of
the ODE) which we label $f^{(s_i)}$ with $i=1 \ldots m$.  Then
construct the $n\times n$ matrix
\begin{equation}
  U = \left( f^{(s_1)} ,\, f^{(s_2)} ,\, \ldots ,\, f^{(s_m)} ,\, a^{(m+1)} ,\, \ldots ,\, a^{(n)} \right) \:,
\end{equation}
where the $a^{(j)}$, $j=m+1 \ldots n$, are linearly independent
constant vectors.  They can be freely chosen, so long as
$\det U \ne 0$ for all $y$.  Usually the $a^{(j)}$ can be unit
vectors.  Now change variables to $g$ by the definition
$f = U \cdot g$ and the equation $f' = A \cdot f$ becomes
\begin{equation}
  \label{eq:g-ode}
  g' = U^{-1} \cdot A \cdot \left( 0,\, \ldots,\, 0,\, a^{(m+1)},\, \ldots,\, a^{(n)} \right) \cdot g \:.
\end{equation}
The first $m$ components of the vector $g$ do not appear on the
right-hand-side of this ODE (after multiplying the matrices out),
so this procedure decouples $m$ of the equations.  Call the
solutions to equation~\eqref{eq:g-ode} $g^{(s_i)}$.  We know $m$
of these are just constant vectors:
$g^{(s_1)}=(1,\,0,\,\ldots,\,0)$,
$g^{(s_2)}=(0,\,1,\,0,\,\ldots,\,0)$ and so on to $g^{(s_m)}$.
The remaining $n-m$ solutions are to be determined using other
techniques.  Once they are found, the remaining solutions to the
original ODE are given by $f^{(s_j)} = U \cdot g^{(s_j)}$.



\end{document}